\documentclass[bibyear]{aa}  
\usepackage{graphicx}
\usepackage{txfonts}
\usepackage{enumerate}
\usepackage{txfonts}
\usepackage{graphicx}

\begin{document}

\title{Learning about stars from their colors} 
  
\author{C. Allende Prieto\inst{1,2}}

   \institute{Instituto de Astrof\'{\i}sica de Canarias,
              V\'{\i}a L\'actea, 38205 La Laguna, Tenerife, Spain\\              
         \and
             Universidad de La Laguna, Departamento de Astrof\'{\i}sica, 
             38206 La Laguna, Tenerife, Spain
             }

%                        \email{callende@iac.es}
%
       %      \thanks{The university of heaven temporarily does not
       %              accept e-mails}

   \date{submitted April 25, 2016; accepted September 23, 2016}

% \abstract{}{}{}{}{} 
% 5 {} token are mandatory
 
  \abstract
   {}
   {We pose the question of how much information on the atmospheric parameters
   of late-type stars can be retrieved purely from color information 
   using standard photometric systems.}  
   {We carried out numerical experiments using stellar fluxes from model atmospheres, injecting random 
   noise before analyzing them. We examined the presence of degeneracies among
   atmospheric parameters, and evaluated 
   how well the parameters are extracted depending on the number and wavelength span 
   of the photometric
   filters available, from the UV GALEX to the mid-IR WISE passbands. We also considered  
   spectrophotometry from the Gaia mission.}
   {We find that stellar effective temperatures can be determined accurately 
   ($\sigma \sim$ 0.01 dex or about 150 K) when reddening 
   is negligible or known, based merely on optical photometry, and the accuracy can be improved twofold
   by including IR data. On the other hand, stellar metallicities and surface gravities are fairly
   unconstrained from optical or IR photometry: $\sim 1$ dex for both parameters at low metallicity, 
   and $\sim 0.5$ dex for [Fe/H] and $\sim 1$ dex for $\log g$ at high metallicity. However, 
   our ability to retrieve these parameters can improve significantly by adding UV photometry. When 
   reddening is considered a free parameter, assuming it can be modeled perfectly, our experiments 
   suggest that it can be disentangled from the rest of the parameters.}
   {This theoretical study indicates that combining broad-band photometry from the UV to the 
   mid-IR allows  atmospheric parameters and interstellar extinction to be determined with 
   fair accuracy, and that the results are moderately robust to the presence of systematic 
   imperfections in our models of stellar spectral energy distributions. The use of UV 
   passbands helps substantially to derive metallicities (down to [Fe/H] $\sim -3$) and surface 
   gravities, as well as to break the degeneracy between effective temperature and reddening. 
   The Gaia BP/RP data can disentangle all the parameters, provided the stellar SEDs are 
   modeled reasonably well.}

   \keywords{techniques: photometric -- stars: atmospheres, fundamental parameters -- 
                dust, extinction
               }

   \maketitle
%
%________________________________________________________________

\section{Introduction}
\label{Intro}

The border between photometry and spectroscopy is poorly defined.
As soon as two photometric measurements, a {\it color}, are available, 
photometry can be considered as very-low resolution  spectroscopy. 
Multicolor photometric systems are now evolving from a few 
 passbands to dozens of  contiguous 
narrow windows (see, e.g., Aparicio Villegas et al. 2010). Obviously, narrower and 
more numerous passbands  covering a wider spectral range are desirable, 
but the information content does not necessarily increase linearly 
with the number of filters.

Bailer-Jones (2004) performed an optimization exercise varying 
the number, width, and location of the passbands to maximize the information 
content for stellar sources and minimize observing efforts. Introducing a new
system is a luxury that  only  very large projects can afford, 
while in most cases data are available on one of a number of widely used 
photometric systems (Bessell 2005).
%It has been pointed
%out that there are fundamental differences in the calibration of photometric
%and spectroscopic observations that cannot be ignored, 
%and therefore the choice 
%between using photometric filters or dispersion elements, such as gratings or
%prisms, is not a trivial one (REF).

\begin{figure*}[t!]
\centering
{\includegraphics[width=16cm]{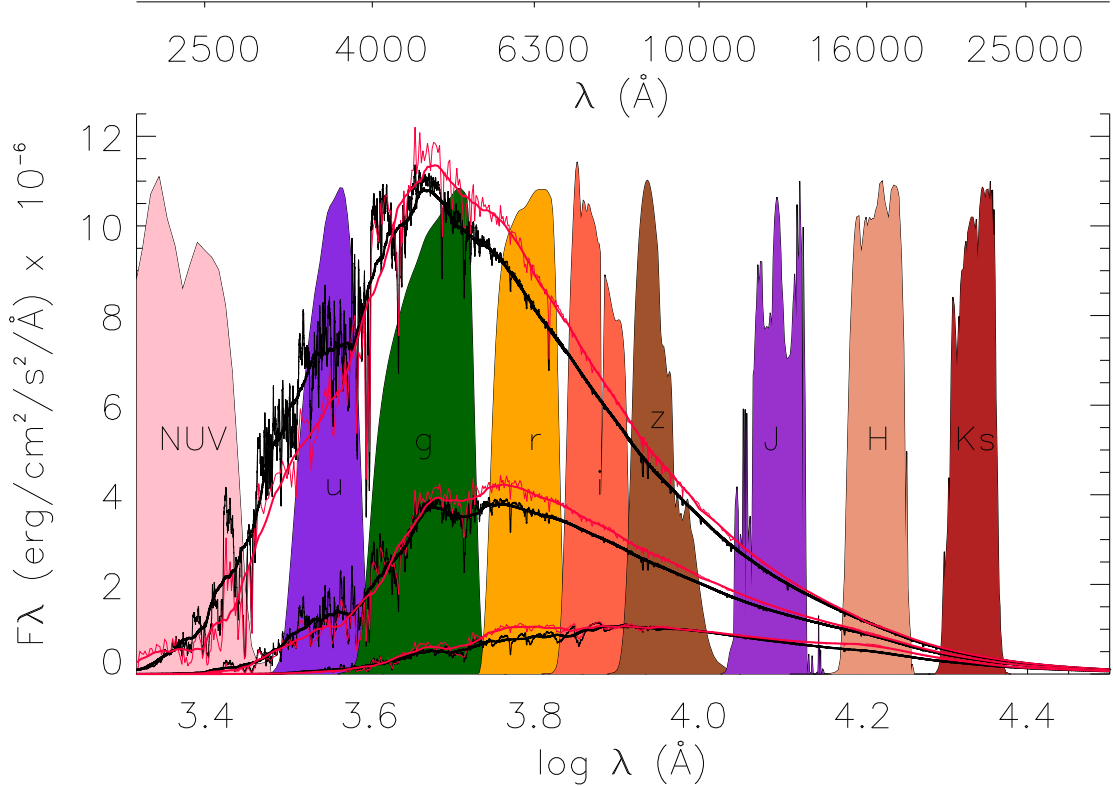}}
\caption{Three sample spectra with solar metallicity, surface gravity $\log g=4$,  
and effective temperatures (in decreasing order of flux level) of 6000, 5000, and 4000 K.
For each model there are two curves, one corresponding to the Kurucz (1993) models (in 
black) and one to the more recent spectral energy distributions by Allende Prieto et al. (2016;
red). Both are shown at a resolving power of $R\sim 200$ and smoothed (thick lines) to $R\sim 8$, 
corresponding to the effective resolution of a broad-band photometric filter. 
The position and shape of the responses for the different filters under 
consideration are shown with an arbitrary normalization. 
The GALEX FUV and the WISE  W1 and W2 bandpasses fall outside of the 
spectral range represented.}
\label{flux}
\end{figure*}

In this paper we make an attempt to quantify the information that can
be extracted from photometric or spectrophotometric observations of stars
using several existing systems. We  focus on some of the most useful ones 
 with wide sky coverage, such as the SDSS $ugriz$ system (Fukugita et 
al. 1996), 2MASS $JHK_s$   (Skrutskie et al. 2006), WISE
(Wright et al. 2010), and the GALEX passbands 
(Martin et al. 2005) as representatives of the optical, near-IR, mid-IR, 
and UV spectral windows, respectively (see Fig. \ref{flux}). 
We examine as well the potential of the Gaia BP/RP spectrophotometry (Jordi 
et al. 2010). 

While substantial work has already been devoted to studying the potential
of many of these (spectro-)photometric systems in order to constrain
stellar parameters and interstellar extinction  (see, e.g., 
Lenz et al. 1998, Masana et al. 2006, Bailer-Jones et al. 2013),
we provide here -- for the first time -- an attempt to make a fair 
comparison of all these systems and the incremental improvement
that may be expected as different spectral windows are
added.

This is a purely theoretical study in which random and systematic
errors are considered by means of simulations. Section \ref{model} describes 
the model spectra used in our analyses and the 
approximations adopted for including interstellar extinction.
Section \ref{method} describes our analysis methodology and figure of merit.
Section \ref{nored} is devoted to the case in which interstellar extinction is
negligible, while Section \ref{red} takes that effect into account.
 Section \ref{xp} considers the important case of 
the spectrophotometric data that will soon become available from the ESA
mission Gaia (de Bruijne 2012). Section \ref{sys} considers the impact
of systematic errors in models, and
Section \ref{summary} summarizes our results and closes the paper.

\section{Model}
\label{model}

As mentioned in the Introduction, this study seeks to quantify 
in an approximate way the potential for extracting information on 
a star's atmospheric parameters from photometry or very-low
dispersion  spectroscopy. To achieve this goal we use
synthetic stellar spectra computed from standard model atmospheres,
smooth them with a wide kernel, sample them at the wavelengths
of the photometric systems of interest, add noise to them, and
finally attempt to recover the atmospheric parameters from 
the simulated photometry.

We want to explore a wide range in the main atmospheric parameters, keeping
the number of variables small enough to be practical. In that spirit,
we work with the stellar effective temperature $T_{\rm eff}$, 
surface gravity $\log g$\footnote{Here $g$
is the gravitational acceleration
at the surface of the star in units of cm s$^{-2}$;  therefore,
$g=G M/R^2$, where $G$ is the gravitational
constant, $M$ is the stellar mass, and $R$ is the stellar radius.}, 
and the overall abundance of metals relative to hydrogen
([Fe/H]\footnote{We use the standard notation [X/Y] 
$= \log \frac{\rm N(X)}{\rm N(Y)} - 
\log \left( \frac{\rm N(X)}{\rm N(Y)}\right)_{\odot}$, where N(X) represents
the number density of nuclei of the element X.}), 
ignoring the fact that abundance ratios
[X/Fe] may be different from 0 (i.e., non-solar) for some metals X.

Spectral energy distributions sampled fine enough for our purpose
are part of the output of standard model atmosphere calculations
with the ATLAS9 code (Kurucz 1979, 1993; Castelli \& Kurucz 2003; 
M\'esz\'aros et al. 2012). The different versions 
of ATLAS9 models available over the years correspond to a sequence
of improvements regarding input atomic and molecular data, 
reference abundances, and algorithms. Since in our tests 
we add noise to the models to simulate observations, and
subsequently attempt to recover the atmospheric parameters,  
it is largely irrelevant which version of the models we adopt. 
We chose the Kurucz (1993) models over more recent versions since
they provide a more complete and regular coverage of late-type stars at
very low metallicities, an area of the parameter space of high interest. 

Kurucz's ATLAS9 provides approximate emergent 
spectral energy distributions similar to those internally used in the code to
evaluate the energy balance\footnote{Accessible from http://kurucz.harvard.edu}. 
They are coarsely sampled, in 
steps that vary between a few angstr\"oms at the UV end 
(starting at 90 \AA) to  20 $\mu$m at the IR limit (160 $\mu$m).
In velocity space the steps are  larger in the UV and optical
range, at about 4000-9000 km s$^{-1}$, than in the near- and mid-infrared 
(1--100 $\mu$m), where they range approximately between 1000 and 2000 km s$^{-1}$).

In addition to this model data set,  in some of our tests we use
a more recent set of synthetic spectra based on a different batch
of ATLAS9 model atmospheres (M\'esz\'aros et al. 2012). These 
are computed with newer line and continuum opacities, different
reference zero-point solar abundances, and a different radiative
transfer code. The details are not terribly relevant for our purposes, 
but are described in Allende Prieto et al. (2016). The point to stress
is that there are important differences between the synthetic
spectral energy distributions described above and computed 
by Kurucz (1993) and these newer ones, and this allows us to estimate
the impact of systematic errors in the models.
Figure \ref{flux} illustrates these differences for three 
models with $T_{\rm eff}=6000$, 5000, and 4000 K, in all cases at
[Fe/H]=0 and $\log g=4.0$.

In some of our experiments we  consider the effect of interstellar
absorption on the observed spectra. We model reddening following  
Fitzpatrick (1999; see also Fitzpatrick \& Masa 1990 and
references therein). We use their software, adopting the average
Galactic value for the parameter $R\equiv A_v/E(B-V) =3.1$.

\section{Methodology}
\label{method}

Stellar atmospheric parameters can be inferred from observations
by adopting a variety of  strategies, depending on spectral type 
and the available data. A recent overview of techniques has been
given by Allende Prieto (2016), and further details can be found 
in Chapters 14--16 of the textbook by Gray (2008). In this paper
we are concerned with the information content of a number of
important photometric or spectrophotometric systems, and it is
therefore appropriate to consider the case in which such
(spectro-)photometric  measurements are the only data available
for the targets of interest. 

In order to gain an understanding of the relative impact of
random and systematic errors, we consider separately the
cases in which only the former or both are included. We  also deal 
with the case where interstellar absorption (reddening) 
is not present, and  study later  its impact on the determination
of stellar parameters. Even though reddening is always present, 
there are cases where it may be negligibly small (for stars 
in the immediate solar neighborhood, at a few parsecs), or where it
is accurately known from other measurements, e.g.,  from IR dust 
emission maps or diffuse bands observed in spectra towards
the same direction).  

We compiled the Kurucz (1993) spectral energy distributions (SED)
with metallicities between $-4.5$ and $+0.5$ (with 0.5 dex steps), 
effective temperatures in the range 4000 to 7000 K (250 K steps), 
and surface gravities spanning  between 1.5 and 4.0 (0.5 dex steps).
We smoothed the SEDs to mimic the spectral resolution  typical of 
broad-band photometric observations 
($R \equiv \lambda/\delta\lambda  \sim 8$;  see the smoothed black
curves in Fig. \ref{flux}) or the Gaia spectrophotometers
($R \sim 30$).  We then interpolated in the regular grid of SEDs using cubic
B\'ezier splines to get SEDs for a random set of parameters uniformly
distributed over the grid. We finally added random noise to these
interpolated fluxes at a level that can be reasonably expected
for real observations:
\begin{itemize} 
\item 10\% (0.1 mag) and 7 \% (0.07 mag) for the GALEX FUV and NUV passbands, respectively;
\item 5\% for the SDSS $u$ and $z$ bands, and 2\% for the other SDSS filters;
\item 4\% for the 2MASS $JHK_s$ photometry; 
\item 4\% for the WISE $W1$ and $W2$ bandpasses.
\end{itemize}

\noindent In the case of the Gaia BP/RP data, we adopted two values for the 
signal-to-noise ratio, 50 and 20 ($\sigma = 2$\% and 5\%), 
independent of wavelength, which are approximately 
the mean values expected at blue wavelengths 
for stars with $G\simeq$ 16 and 18 (Bailer-Jones 2010).

The resulting SEDs are analyzed using the optimization code
FERRE\footnote{FERRE is available from http://hebe.as.utexas.edu/ferre} 
(Allende Prieto et al. 2006), 
which infers the atmospheric parameters from the very same models used
to simulate observations. FERRE couples an optimization algorithm,
in this case Powell's UOBYQA algorithm (Powell 2000), with interpolation 
in a grid of model spectra. In this analysis we used linear interpolation
to avoid adopting exactly the same interpolation scheme used to prepare
the simulations. 

These tests are ideal in the sense that no systematic errors other than
those incurred by adopting linear interpolation in FERRE are considered.
However, they are useful to reveal potential degeneracies among the parameters
regarding the predicted SEDs. The fact that we have introduced random
noise in the simulated fluxes  hampers our ability to recover the 
true underlying model parameters. We examine in particular  the effect
of including different bandpasses on the recovery of the atmospheric parameters.

\section{Photometric tests without reddening}
\label{nored}

We considered first the simplified situation in which there is no reddening
distorting the observed stellar spectra. We studied five combinations of
photometric systems, corresponding to the  available data:   
\begin{itemize}
\item  only SDSS $ugriz$ photometry; 
\item   SDSS and 2MASS $JHK_s$ photometry;
\item  SDSS, 2MASS, and GALEX $FUV$ and $NUV$ photometry;
\item  SDSS, 2MASS, and WISE $W1$ and $W2$ photometry;
\item  SDSS, 2MASS, GALEX, and WISE  photometry.
\end{itemize}

\begin{figure*}
\centering
{\includegraphics[angle=90,width=16cm]{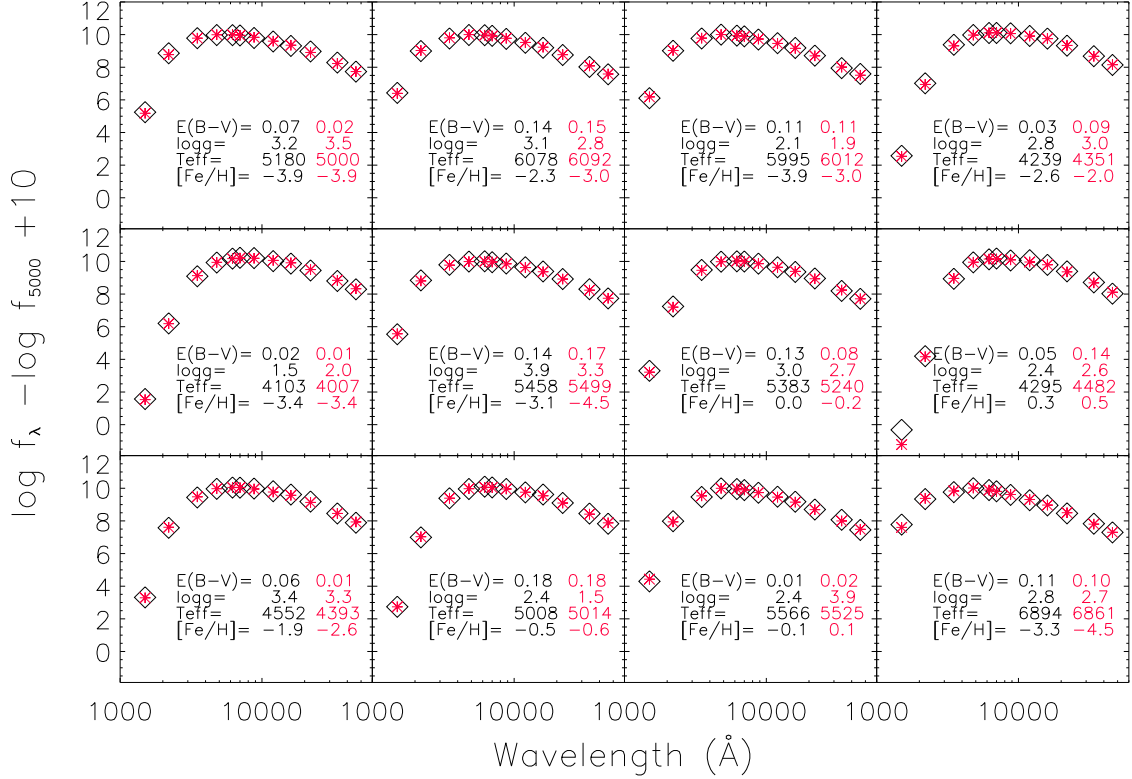}}
\caption{Comparison between the simulated photometric observations (open symbols) 
and those recovered from the fittings (red asterisks) for a few test cases 
chosen at random.
The labels show the parameters 
recovered with FERRE (red) 
and the {true} values (black).}
\label{mdl}
\end{figure*}

We ran the simulated data, transformed into normalized logarithmic units, 
through the fitting code and recovered the parameters for each object.
Fig. \ref{mdl} illustrates the input and the output data for one of
our experiments with reddening;  except for the presence of that
parameter, those without reddening are very similar.

For each case we calculate the dispersion between
the recovered and the {true} parameters by finding the width of the 
distribution excluding 15.85\% of the data at each extreme (those with 
the largest discrepancies) and dividing that value by two. For a Gaussian
distribution this value would correspond to 1$\sigma$, but this procedure
makes the determination robust to a small percentage of outliers.
We do this for the full sample of 1000 simulated targets, and only for
the roughly 300 that lie at metallicities higher than [Fe/H]$=-1$. 
This separation is useful
since most stars in the Milky Way reside in the Galactic disk
in that metallicity range, and at high metallicities there is significantly
more information than at low values owing to the presence of numerous and strong
absorption lines in the spectra. In fact, there is nearly no information
on metallicity in spectra at [Fe/H]$<-3$.

\begin{table*}
\centering
\begin{tabular}{lcccccc}
\hline
      &   \multicolumn{3}{c}{$-5\le$ [Fe/H]$\le +1$} &  \multicolumn{3}{c}{$-1 \le $[Fe/H]$\le +1$} \\      
\hline
          bandpasses  & [Fe/H] & $T_{\rm eff}$ & $\log g$ & [Fe/H] & $T_{\rm eff}$ & $\log g$ \\
                      &          & (K)  & (cgs) &      & (K)  & (cgs) \\
\hline
\hline
                         $ugriz$ &   1.2 & 129  &   1.0  &  0.5 & 148  &   0.9 \\
                    ugriz JHKs &   1.2 &  62  &   0.9  &  0.3 &  51  &   0.8 \\
            F-NUV ugriz  JHKs &   0.7 &  53  &   0.6  &  0.2 &  52  &   0.5 \\
               ugriz JHKs W1W2 &   1.3 &  56  &   1.1  &  0.4 &  46  &   1.0 \\
        FUVNUV ugriz JHKs W1W2 &    0.7 &  45  &   0.6 &   0.2 &  44  &   0.5 \\               
\hline
                   BP/RP (S/N=50) &   0.2  & 23 &   0.2  &  0.1  & 27   &  0.2 \\
                   BP/RP (S/N=20) &   0.5  & 57 &   0.4  &  0.2  & 61   &  0.4 \\
\hline
\end{tabular}
\caption{Robust standard deviation between the retrieved and {true} atmospheric parameters 
for the full sample ($-5\le$[Fe/H]$\le+1$) and a metal-rich subsample ([Fe/H]$\ge -1$) as a function
of the photometric bandpasses included in the numerical experiments. No reddening is included.}
\label{t1}
\end{table*}

\begin{figure*}
\centering
{\includegraphics[angle=90,width=16cm]{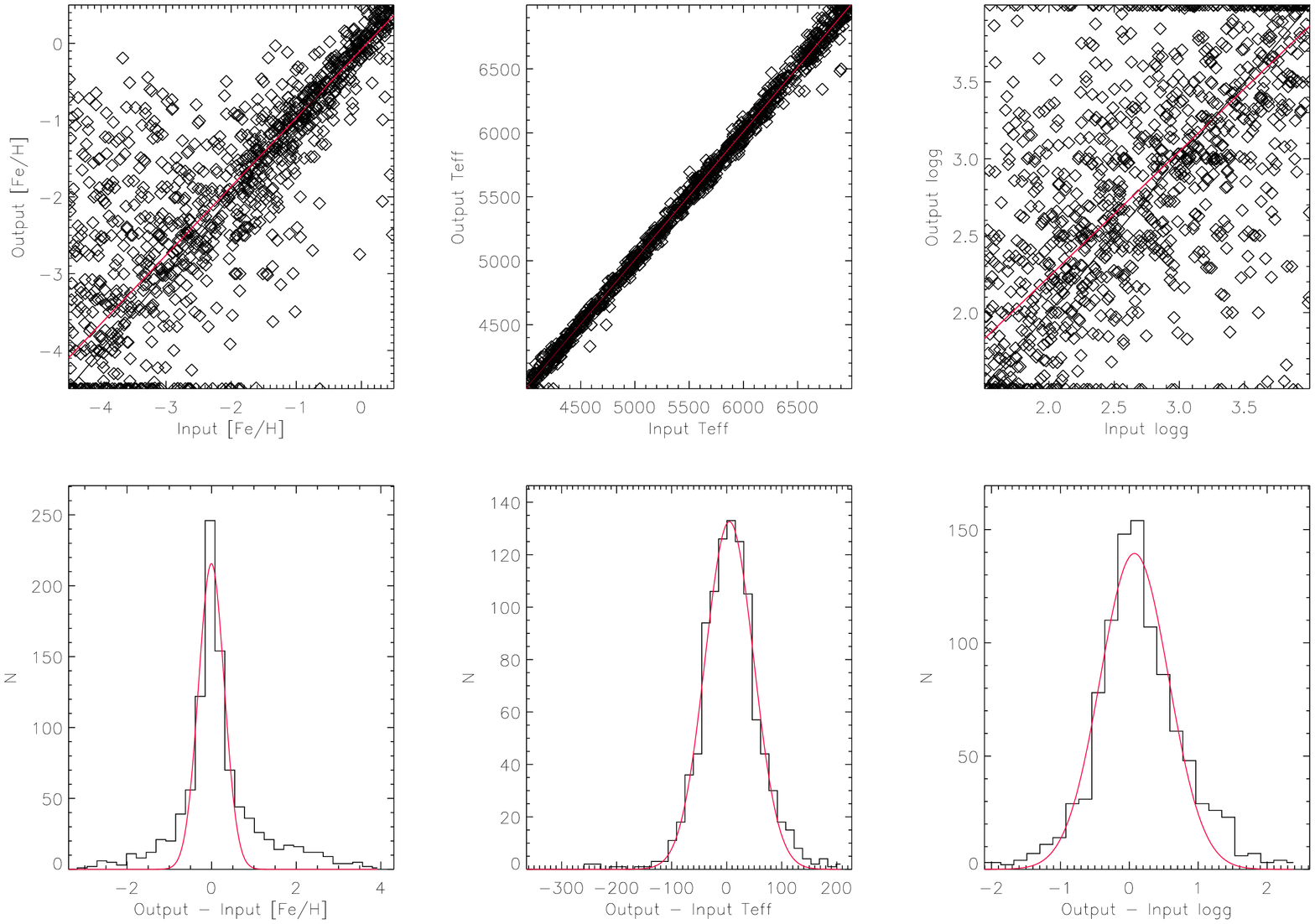}}
\caption{Upper panels: Comparison between the parameters recovered with FERRE from
our simulations against the {true} values for the case when all the filters are
included. The red curves show a least-squares fit
to the data. Bottom panels: Histograms of the residuals
for each parameter. The red curves correspond to a Gaussian fit to the data, which 
systematically underestimates the width of the distribution for [Fe/H]; therefore, it is
not adopted for the statistics shown in the tables and is discussed in the text. The median 
offsets of the distributions are 0.03 dex, 6 K, and 0.08 dex for [Fe/H], $T_{\rm eff}$, 
and $\log g$, respectively. The straight standard deviations are 0.98 dex, 58 K, and 
0.67 dex,  while those computed with the robust algorithm described 
in the text are 0.70 dex, 45 K, and 0.57 dex.}
\label{cmp}
\end{figure*}

The results are given in Table \ref{t1}, and Fig. \ref{cmp} illustrates
the analysis for the full sample in the case when all photometric bandpasses
are considered.  
The uncertainties in all three parameters vary in sync with  
larger errors for more metal-poor stars, a more modest dependence on 
effective temperature (in the narrow range where are experiments are
contained), and a fairly weak dependence on surface gravity.
We find that the average
value of the  distributions have very small offsets from zero, and therefore 
the table gives only the robust standard deviation for the distribution
of residuals in each parameter. Our statistics are slightly distorted by the
fact that solutions cannot be found outside the limits of our model grid, but
these distortions appear to be modest.

SDSS photometry ($ugriz$) alone can constrain fairly well 
the stellar effective temperature ($\sigma \sim 150$ K or about 0.01 dex), 
but poorly constrains the metallicities or stellar surface gravities. Adding infrared
photometry significantly improves the recovery of $T_{\rm eff}$, but has very
little impact on gravity, while the results for [Fe/H] vary depending on 
the metallicity range:  at high metallicity, optical and near-infrared photometry
help in recovering this parameter, but offer little leverage for more metal-poor stars. 
Adding UV photometry can  significantly improve the recovery of metallicities 
and surface gravities for stars at all metallicities.

\section{Photometric tests including reddening}
\label{red}

In most practical situations interstellar reddening gets in the way, making
it more difficult to derive  the parameter that has the highest influence on the shape
of the spectrum, the effective temperature. Naturally, the determination of
the other two atmospheric parameters will be severely affected by the 
potential confusion between reddening and $T_{\rm eff}$.

We  repeated the experiments described in Section \ref{nored} adding
an additional dimension to the problem, accounting for the effect of reddening
as described in \S \ref{model}. Each of the 1000 simulated observations were reddened
for a random value of $E(B-V)$ between 0 and 0.25 mag drawn from a uniform 
distribution. This range is typical outside a stripe of $\pm 10 \deg$ from the 
Galactic plane.

\begin{table*}
\centering
\begin{tabular}{lcccccccc}
\hline
      &   \multicolumn{4}{c}{$-5\le$[Fe/H]$ \le 1$} &  \multicolumn{4}{c}{$-1 \le$[Fe/H]$\le +1$} \\      
\hline
         bandpasses & [Fe/H] & $T_{\rm eff}$ & $\log g$ & E(B-V) & [Fe/H] & $T_{\rm eff}$ & $\log g$ & E(B-V) \\
                               &          & (K)  & (cgs) & (mag) &       & (K)  & (cgs)& (mag) \\
\hline
\hline
                         $ugriz$    & 1.3 & 332 &    0.9 &  0.08 &   0.6 & 372  &   1.1 &  0.08 \\
                    $ugriz$ $JHKs$    & 1.1 & 275 &    0.9 &  0.07 &   0.3 & 242  &   0.8 &  0.07 \\
            $F-NUV$ $ugriz$  $JHKs$    & 0.9 & 134 &    0.6 &  0.04 &   0.3 & 151  &   0.6 &  0.04 \\
               $ugriz$ $JHKs$ $W1W2$    & 1.2 & 265 &    1.0 &  0.07 &   0.4 & 224  &   1.0 &  0.07 \\
        $F-NUV$ $ugriz$ $JHKs$ $W1W2$    & 0.9 & 137 &    0.6 &  0.04 &   0.3 & 136  &   0.5 &  0.04 \\
\hline
                  BP/RP (S/N=50)  &  0.4 & 153  &   0.3 &  0.03  &  0.2 & 125   &  0.3  & 0.03 \\
                  BP/RP (S/N=20)  &  0.7 & 239  &   0.5 &  0.05  &  0.3 & 236   &  0.5  & 0.05 \\
\hline
\end{tabular}
\caption{Same as Table \ref{t1}, but for the case in which interstellar reddening is included in
the experiments.}
\label{t2}
\end{table*}

Similarly to the case in previous section, we used FERRE to recover the
atmospheric parameters and the reddening $E(B-V)$ simultaneously (See Fig. \ref{mdl}). 
The  results in this case are summarized in Table \ref{t2}. Obviously, our ability 
to recover  all the atmospheric parameters is significantly degraded compared to 
zero-reddening case presented in Section \ref{nored}. Nonetheless, qualitatively, most of our  
conclusions hold: effective temperatures are already recovered 
with optical photometry at a level of about 350 K (0.03 dex), and this value improves 
somewhat adding near-infrared photometry from 2MASS. We now find that our ability 
to discern the effects of reddening from those of interstellar extinction is substantially
improved by folding UV photometry in the data set, and this has a positive impact on
all parameters. Finally, adding the mid-IR data from WISE adds very little information.

\section{Gaia spectrophotometry}
\label{xp}

Gaia will provide spectral energy distributions for about $10^9$ stars in the Milky Way. 
Operating in space, it is expected that these data will be very precise, even though the 
zero-point accuracy will likely be limited by difficulties  connecting
standard flux laboratory sources and those made with an astronomical instrument
(Bohlin et al. 2014). 

We repeated the experiments in the previous sections at somewhat 
higher resolution ($R\sim 25$), sampling the spectra between 3600 and 10500 \AA\
with 86 samples evenly distributed in $\log \lambda$. Such a setup resembles,
but  is not equal to, the data expected from the Gaia photometers BP/RP.
The real data will have a resolving power and sampling
rate that varies with wavelength and shows an abrupt discontinuity
at about 660 nm where the two overlapping blue (BP) and red (RP) channels
meet (see de Bruijne et al. 2014 for more details). In addition, we injected noise 
at a constant level at all wavelengths (5\% and 2\%; see Section \ref{model}),
while the actual observations have a signal-to-noise ratio that depends strongly on 
wavelength (see, e.g., Bailer-Jones 2010). We  also assume that the instrumental
response is perfectly known (see Jordi 2015 for real BP/RP images and collapsed spectra).
Nonetheless, we expect that 
the impact on our conclusions of simulating the Gaia data in more detail  
will be modest, especially in comparison with some of our other approximations.

%                                 -5<=[Fe/H]<=1         -1<=[Fe/H]<=+1      
%                               -------------------------------------------
%                bandpasses      [Fe/H]  Teff  logg  [Fe/H]  Teff  logg
%                  s/n=50 BP/RP    0.2   23     0.2    0.1   27     0.2
%                  s/n=20 BP/RP    0.5   57     0.4    0.2   61     0.4

As shown in Table \ref{t1}, for the Gaia BP/RP spectrophotometry alone, and without extinction 
in our simulations, we find that we are able to recover the stellar metallicities, 
effective temperatures, and 
surface gravities to within 0.2 dex, 23 K, and 0.2 dex, respectively, for a star with
$G\sim 16$ ($S/N \sim 50$), and within 0.5 dex, 57 K, and 0.4 dex, respectively, at
$G\sim 18$ ($S/N \sim 20$).
Limiting the sample to stars more metal-rich than [Fe/H]$=-1$, the uncertainty in [Fe/H] 
improves to 0.1 dex at $G\sim 16$ and 0.2 dex at $G\sim 18$. These figures are 
in fact quite similar to those reported by Bailer-Jones (2010).

%                                 -5<=[Fe/H]<=1         -1<=[Fe/H]<=+1      
%                               -------------------------------------------
%                bandpasses      [Fe/H]  Teff  logg  E(B-V) [Fe/H]  Teff  logg E(B-V)
%              s/n=50     BP/RP    0.4  153     0.3   0.03    0.2  125     0.3   0.03
%              s/n=20     BP/RP    0.7  239     0.5   0.05    0.3  236     0.5   0.05

As we concluded in the case of photometry, there is increased confusion when
reddening is considered (see Table \ref{t2}). The uncertainties in the recovered values of 
[Fe/H], $T_{\rm eff}$, and  $\log g$ rise to 0.4 dex, 153 K, and 0.3 dex, respectively, 
for the full sample, or 0.2 dex, 125 K, and 0.3 dex for the subsample at [Fe/H]$>-1$.
In both cases the reddening is recovered with an uncertainty of about 0.03 mag.

\section{Systematic errors}
\label{sys}

It is interesting to explore the impact of systematic errors on the 
model stellar fluxes. We can get a glimpse of how much such effects
can distort our results by analyzing the simulated photometry, based on 
the Kurucz (1993) SEDs described above, with a different set of models
by Allende Prieto et al. (in preparation).

The fluxes computed  by Allende Prieto et al. (in prep.) are more limited
in wavelength coverage (they do not cover in full the GALEX passbands) and parameter
space than the Kurucz (1993) SEDs, as described in \S \ref{model}. For that
reason we limit our tests to the optical and infrared filters, and
to stars with $T_{\rm eff}< 6000$ K and [Fe/H]$>-1$. The limit on 
$T_{\rm eff}$ is not strict since the new model fluxes are available
for $T_{\rm eff}>6000$ K, but warmer models are part of a separate bundle
from those for cooler temperatures, and including stars between 6000 and
7000 K would significantly complicate the implementation of our tests.

%                                 -2.5<=[Fe/H]<=1         -1<=[Fe/H]<=the+1      
%                               -------------------------------------------
%                bandpasses      [Fe/H]  Teff  logg  [Fe/H]  Teff  logg
%                         ugriz    0.7  141     1.9    0.5  145     1.9
%                    ugriz JHKs    0.6   68     1.7    0.4   64     1.6
%               ugriz JHKs W1W2    0.7   65     2.0    0.4   65     1.9

\begin{table*}
\centering
\begin{tabular}{lcccccc}
\hline
      &    \multicolumn{6}{c}{$-1 \le $[Fe/H]$\le +1$} \\      
\hline
      &    \multicolumn{3}{c}{median offset}  &    \multicolumn{3}{c}{robust $\sigma$} \\
\hline
          bandpasses  & [Fe/H] & $T_{\rm eff}$ & $\log g$ & [Fe/H] & $T_{\rm eff}$ & $\log g$ \\
                      &          & (K)  & (cgs) &      & (K)  & (cgs) \\
\hline
\hline
                         $ugriz$ &  0.0 & -28  &  -0.3 &   0.5 & 145  &   1.9 \\
                    ugriz JHKs &   0.3  & 12  &   0.4  &  0.4 &  64  &   1.6 \\
               ugriz JHKs W1W2 &   0.2 &  33  &  -0.1  &  0.4 &  65  &   1.9 \\
\hline
\end{tabular}
\caption{Median offsets and robust standard deviations between the retrieved and {true} 
atmospheric parameters for a metal-rich subsample ([Fe/H]$\ge -1$). No reddening is considered,
 similar to the results reported in Table \ref{t1}, but systematics errors in modeling the
 spectra are included by analyzing the simulations with an inconsistent set of models.}
\label{t3}
\end{table*}

The results in this case are shown in Table \ref{t3}. Unlike Tables \ref{t1} and \ref{t2}, this
table includes the median offset between the parameters recovered by FERRE and the
{true} ones in the first three columns in addition to the robust standard
deviation in columns $4-6$, which can be directly compared to those in Table \ref{t1}
for the case limited to $-1 \le $[Fe/H]$\le +1$ (although it corresponds to a slightly
smaller range in $T_{\rm eff}$). It should be remembered that the median offsets of the
distribution of residuals from zero were negligible when systematic errors were not
considered in Sections \ref{nored}, \ref{red}, and \ref{xp}.

Overall, the retrieved parameters suffer modest systematic errors compared to the 
spread  even though the differences between the models used in the simulations
and the analysis are important (see examples in Fig. \ref{flux}). 
The widths of the distributions are very similar to the values reported
in Table \ref{t1} for [Fe/H] and $T_{\rm eff}$, but the surface 
gravity  recovered is significantly worse when systematic errors are taken into account.

\section{Summary and conclusions}
\label{summary}

We perform numerical experiments to explore the potential of broad-band photometry
and spectrophotometry of stars to infer atmospheric parameters and the 
interstellar extinction along the line of sight. Our experiments are fairly simplistic
in that we ignore systematic errors that result from an imperfect knowledge of the
distortions introduced by instruments and, in the case of ground-based observations,
the effect of Earth's atmosphere, but should hold as long as they are significantly
smaller than the random uncertainties adopted (2-10\%, depending on the instrument
and wavelength). As such, our results provide an upper limit to the performance 
achievable for real data.

We find that photometry over a wide spectral range from the near-UV to the mid-IR
can constrain  the atmospheric parameters fairly well. While optical data are
enough to constrain well the effective temperature of late-type stars, this
parameter is retrieved with far better precision by including infrared observations.
We show that the determination of surface gravity and metallicity can benefit 
substantially from the addition of near-UV photometry. 

For a similar signal-to-noise ratio, the performance expected for an instrument 
like the Gaia BP/RP spectrophotometer is better than standard photometric systems
with a similar wavelength span. This is not surprising, since both resolution and sampling
are much better for BP/RP than for the standard broad-band photometric systems. However,
for the faintest stars observed by Gaia, photon noise will limit significantly the 
atmospheric parameters 
determined from the BP/RP observations.

Photometric searches for ultra-metal-poor stars can benefit enormously from
counting on UV passbands. In our tests without reddening, the success rate 
at [Fe/H]$\le -3$, defined as the fraction of stars correctly identified
in that range, and the false-positive rate are 59\% and 42\%, respectively,
using only SDSS photometry, but these figures change to 79\% and 22\%, respectively,
when the GALEX filters are included in the observations. 

The effect of interstellar absorption on the observations hampers our 
ability to recover the stellar atmospheric parameters. Only with wide wavelength
coverage -- in particular in the blue and into the near-UV -- and high signal-to-noise
ratios can reddening be cleanly disentangled from variations in 
the atmospheric parameters. 

It will be very interesting to verify our expectations with real data, although it 
goes beyond  the scope of the present paper and will likely require some sort
of zero-point calibration of the synthetic photometry. In addition, 
in our numerical experiments we have sampled the spectral energy distributions 
at specific wavelengths corresponding to each of the bandpasses under study, 
but for practical applications it is necessary to convolve the model SEDs 
with the filter responses since a filter's effective wavelength 
depends on the stellar spectrum.

Our main conclusion is that wide-area photometry at multiple 
wavelengths is a promising path for  characterizing the stellar populations of
the Milky Way and other nearby galaxies where stars can be resolved. 
Following up on pioneering studies (e.g., Ivezi\'c et al. 2008), 
projects that already do this in a homogeneous fashion such as ALHAMBRA (Moles et al. 2008), 
J-PLUS and JPAS (Ben\'{\i}tez et al. 2014), PAU (Castander et al. 2012), 
or the Gaia mission will demonstrate the potential and the limitations of 
this technique. 

\begin{acknowledgements}

I am grateful to David S. Aguado y Jonay Gonz\'alez Hern\'andez
for valuable discussions and constructive criticism on this manuscript.
I also appreciate useful suggestions from the referee, Eduard Masana, and 
the excellent work by Helenka Kinnan improving the language.
My research has been generously supported by the Spanish MINECO 
(AYA2014-56359-P).

\end{acknowledgements}

\end{document}